\documentclass[12pt,a4paper]{article}
\usepackage{a4wide}
\usepackage{latexsym}
\usepackage{epsf}
\usepackage{multicol}

\date{}

\begin{document}

\begin{flushright}

\small

IFT-UAM/CSIC-99-43\\




\end{flushright}


\begin{center}

\vspace{.7cm}
{\Large {\bf A Comment on Fisher Information\\ and Quantum
Algorithms}}\\

\vspace{.7cm}


{\bf\large J.J. Alvarez}${}^{\ddagger}$\footnote{E-mail: {\tt
juanjose.alvarez@uam.es}}
{\bf\large and C. G\'omez}${}^{\diamond\ddagger}$\footnote{E-mail: {\tt
cesar.gomez@uam.es}}

\vskip 0.4truecm

${}^{\ddagger}$\ {\it Instituto de F\'{\i}sica Te\'orica, C-XVI,

Universidad Aut\'onoma de Madrid \\

E-28049-Madrid, Spain}

\vskip 0.2cm

${}^{\diamond}$\ {\it I.M.A.F.F., C.S.I.C., Serrano 113bis\\

E-28006-Madrid, Spain}

\vspace{.7cm}


{\bf Abstract}

\end{center}
\begin{quotation}
\small

We show that Grover's algorithm defines a geodesic in quantum Hilbert
space with the Fubini-Study metric. From statistical point of view
Grover's algorithm is characterized by constant Fisher's function. Quantum
algorithms changing complexity class as Shor's factorization does not
preserve constant Fisher's information. An adiabatic quantum factorization
algorithm in non polynomial time is presented to exemplify the result. 

\end{quotation}
\setlength{\columnsep}{1cm}
\setlength{\columnseprule}{0pt}
\begin{multicols}{2}
  {\bf \large 1.- } Recently a lot of attention has been paid to the
  problem of defining quantum algorithms \cite{REVIEW} \cite{SHOR}
  \cite{GRO}. Generically a quantum algorithm defines a discrete path
  in a quantum Hilbert space with the end point of the path
  corresponding to a quantum state that, after an appropriated
  measurement, will eventually provide, with high probability, the
  answer to a given problem.  In this note we will work out some
  geometrical aspects of quantum algorithms. We will consider first the
example of Grover's algorithm \cite{GRO}. In this case it can be shown
that the path
  - in the quantum Hilbert space- associated with the algorithm, is a
  geodesic in Fubini-Study metric \cite{WOOTTERS}.  Geodesics in
  quantum Hilbert space are intimately connected with Fisher
  information function \cite{BRANDT}. Using Fisher's function we will
define a formal Lagrangian on probability space such that their
trajectories coincide with the quantum algorithm path.
Next we will work out Shor's factorization algorithm. We will show that,
in this case as in any other involving change in complexity class (from NP
to P problem), the Fisher's function does not remain constant. However it
is possible to design a factorization algorithm using Grover's scheme. In
this factorization algorithm Fisher's function remains constant but
``computing time'' is non polynomial (the process is adiabatic with
respect to Fisher's ``entropy''). This strongly indicates that changes in
complexity class require no conservation of Fisher's function - i.e. non
unitarity quantum state projection.\\[.2cm]
  {\bf \large 2.- } Grover's algorithm provides a way to find, by
  means of a quantum computer \cite{REVIEW}, one particular item - in
  a set of N items randomly ordered -after approximately
  $\frac{\pi}{4}\,\sqrt{N}$ iterations. This algorithm is known to be
  optimal \cite{ZALKA}. In order to define the algorithm let us
  introduce the quantum state:
\begin{equation}
\label{initial}
\mid\psi\rangle\,=\,\frac{1}{\sqrt{N}}\sum_{i=0}^{N\,-\,1}\mid i\rangle
\end{equation}
The algorithm is determined by a set of states $\mid\psi\rangle_j$:
\begin{equation}
\label{INITIAL}
\mid\psi\rangle_j\,=\, k_{j}\mid 0\rangle\,+\,\sum_{i\,=\,1}^{N-1}\,
l_{j}\mid i\rangle
\end{equation}
with:
\begin{eqnarray}
\label{recur}
k_{j+1} & = & \frac{N-2}{N}\,k_{j}\, +\, 2 \,\frac{N-1}{N}\,l_{j}
\nonumber\\
l_{j+1} & = & \frac{-2}{N}\,k_{j}\, + \,\frac{N-2}{N}\,l_{j}
\end{eqnarray}
where the state $\mid0\rangle$ represents the item we are looking for
\cite{CANAD}. We will think of (\ref{INITIAL}) as the discrete path
defining Grover's algorithm.
Let us approximate the discrete path (\ref{INITIAL}) by a path:
\begin{equation}
\mid\psi\rangle(\phi)\,=\,\sum_{j=0}^{N\,-\,1}c_{j}(\phi)\,\mid
j\rangle
\end{equation}
depending on a continuous parameter $\phi$. The probabilities
$p_{j}(\phi)\,=\,|c_{j}(\phi)|^2$ to find item $j$ at "computer time"
$\phi$
are - for Grover's algorithm - given by:
\begin{eqnarray}
\label{array}
p_{0}(\phi)&=&\sin^2\phi\nonumber\\
p_{j}(\phi)&=&\frac{\cos^2\phi}{N\,-\,1}\,\,\,\,j\neq0
\end{eqnarray}
These probabilities define a path on "probability space".
Transitions from $\mid\psi\rangle(\phi)$ to
$\mid\psi\rangle(\phi\,+\,\delta\phi)$ are associated with the quantum
computer operations performed by means of quantum gates. If these
transformations are unitary we get:
\begin{equation}
\label{normadot}
\langle\dot{\psi} \mid \dot{\psi}
\rangle\,=\,\frac{1}{4}\,\sum_{j=1}^{N}
\,\frac{\dot{p_{j}}^2}{p_j}\,=\,1
\end{equation}
provided we normalize the state $\mid\psi\rangle$, and where
$\dot{p_j}\,=\,\frac{dp_j}{d\phi}$.
Equation (\ref{normadot}) is our first contact with Fisher's
information
function. In fact defining \cite {BRANDT}:
\begin{equation}
\label{fisher}
{\cal F}(\phi)\,=\, \sum_{i=1}^{N}\frac{\dot{p_{i}}^2}{p_{i}}
\end{equation}
we notice that a path of states generated by unitary transformations
are associated with a one parameter family of probability distributions of
a constant Fisher function of value equal to four.\\[.2cm]
{\bf \large 3.- }Introducing quantum phases by
\\$c_{j}=\sqrt{p_{j}}\,e^{\varphi_{j}}$ the Fubini-Study metric on Hilbert
space is given by:
\begin{equation}
ds_{F-S}^2\,=\,\frac{1}{4}\, \sum_{j=1}^{N}\,\frac{dp_{j}^2}{p_{j}}
\,+\, \left [\sum_{j=1}^{N}\,p_{j}\,d\varphi_{j}^2\,-\, \left
(\sum_{j=1}^{N}\,p_{j}\,d\varphi_{j} \right )^2 \right ].
\end{equation}
The induced metric on a path $(p_{j}(\phi)\,,\,\varphi_{j}(\phi))$ is
given by:
\begin{equation}
ds_{Ind.}^2\,=\, \frac{1}{4}\,\left ({\cal
F}(\phi) \,+\,4\, \sigma_{\dot {\varphi}}^2 \right )\, d\phi^2
\end{equation}
with $\dot {\varphi} \,=\, \frac{d\varphi}{d\phi}$, and ${\cal
F}(\phi)$ the
Fisher function defined in (\ref{fisher}).
For a path with\\ $\dot {\varphi} \,=\,0$ \footnote{This in particular
means that entanglement remains constant}, as the one defined by
Grover's algorithm, the geodesic is given by minimizing \footnote{Use of
Fisher's function to define variational problems has been also considered
in \cite{FISHER}}
\begin{equation}
\label {minimal}
{\cal S}\,=\, \frac{1}{2}\int_A^B\left ({\cal F}_{\phi}\right
)^{1/2}\,d\phi
\end{equation}
with the constraint:
\begin{equation}
\sum_{i=1}^{N}\,p_{i}\,=\,1.
\end{equation}
Defining new variables $p_i=x_{i}^{2}$ the equations of motion for
the
"Lagrangian" $\frac{1}{2}\,\left ({\cal F}_{\phi}\right )^{1/2}$ are:
\begin{equation}
\label{A}
\ddot{x_{i}}\,-\,\left ( \frac{\dot{\cal F_{\phi}}}{{\cal F_{\phi}}}
\right
)\,\dot{x_i}\,+\,\frac{{\cal F_{\phi}}}{4}\,x_i\,=\,0.
\end{equation}
For any quantum algorithm performed by successive unitary
transformations, we
know ${\cal F}(\phi)\,=\,cte.$ reducing (\ref{A}) to the harmonic
oscillator
equation:
\begin{equation}
\label{B}
\ddot{x_{i}}\,+\,\frac{{\cal F_{\phi}}}{4}\,x_i\,=\,0
\end{equation}
where the natural frequency is given by $\omega^2 \,=\, \frac{{\cal
    F_{\phi}}}{4}\,=\,1$. It is now easy to check that Grover's path
(\ref{array}) is in fact solution to (\ref{B}).  Thus, we conclude
that Grover's algorithm defines a geodesic path in
quantum Hilbert space.\\[.2cm]
{\bf \large 4.- }Obviously we can always transform a quantum algorithm of
Grover's type into a one parameter family of probability distributions
$p_{i}(\phi)$
with $i$ running over the Hilbert space basis.  What we have pointed
out in this note, is that this family of probability distributions is
completely determined by unitarity and the condition of minima for
"Fisher's information action" (\ref{minimal}). Notice that in our
definition of Fisher's information function the "computing time"
$\phi$ is playing the statistical role of an statistical estimator. In
particular with respect to this "computing time", ${\cal F}(\phi)$ is
constant as a consequence of unitarity. Hence, in Grover's algorithm,
the "input information" at the starting point $\phi\,=\,\phi_0$ of the
computation is given by:
\begin{equation}\label{input}
\left. \sum_{i=0}^{N\,-\,1}\,\frac{\left (\frac{\partial
P_{i}(\phi)}{\partial \phi}\right )^2}{p_i(\phi)}\right
|_{\phi\,=\,\phi_0}
\end{equation}
and it is this quantity the one that remains constant in the process.
This is in contrast to the evolution of the standard Fisher's
information -contained in $\{ p_{i}(\phi)\}$- concerning where is the
item we are looking for.  Obviously this second form of information
increases in the process until reaching its maximum corresponding to
the point where we find the desired solution. It is interesting to
observe that the "input information" (\ref{input}) is determined by
quantum unitarity and can not be smaller or bigger. In summary, we
conclude that from the information theory point of view quantum
computations of Grover's type appears as equivalent to classical
statistical processes governed by minimum Fisher's action.\\ [.2cm]
{\bf \large 5.- }Next let us consider Shor's factorization algorithm \cite
{SHOR}. As it is well known, classical algorithms for number factorization
require exponential time $\exp (c(\log N)^{1/3}\,(\log \log{N})^{2/3})$
where N is the integer we want to factorize an c is some constant. From
complexity theory, number factorization is considered a NP-problem. Given
a number N we can reduce the problem of factorizing N to find the period
{\it r} of the function $f(a)\,=\,y^a\, mod\,N$ for a random number $y$
smaller than N and coprime with N. In Shor's quantum algorithm the period
of $f(a)$ is obtained in two steps. First we define the quantum register
state:
\begin{equation}
\label{shori}
\mid\Psi\rangle\,=\,\frac{1}{\sqrt{q}}\sum_{a=0}^{q\,-\,1}\mid
a\rangle\,\mid y^a\, mod\,N \rangle
\end{equation}
with $N^2\,<\,q\,<2N^2$. Then, we measure  the value of $\mid y^a\, mod\,N
\rangle$. For each eigenvalue $l$ we get the state:
\begin{equation}
\label{shorf}
\mid
\chi_{l}\rangle\,=\,\sqrt{\frac{r}{q}}\,\sum_{n\,=\,0}^{q/r\,-\,1}\mid
l\,+\,nr\rangle.
\end{equation}
The next step is to proceed by a discrete Fourier transform to wash out
the dependence on $l$. At the end of the process we get the desired period
{\it r} in polynomial time: ${\cal O}((\log N)^2\,(\log \log N)(\log \log
\log N))$.
In this algorithm there are series of unitary transformations we can model
out in terms of standard quantum gates and a typically non unitary process
consisting in the measurement projecting from the register state
(\ref{shori}) to state (\ref{shorf}).
As it is clear from our previous discussion, Fisher information will be
conserved during the unitary discrete Fourier transform but will
generically change in the non unitary measurement process. This change is,
as we will see in a moment, related to the change in complexity class
achieved by Shor's quantum algorithm.
In order to visualize this more clearly let us design a way to find the
period of $f(a)$ using Grover's type of algorithm.
We start with the quantum register state (\ref{shori}). Let us define the
following transformation:
\begin{eqnarray}
\label{rotation}
{\cal C}[\mid a \rangle \,\mid f(a)\rangle]&=&1\,\, if\,\, f(a)\,=\,
f(1)\nonumber \\
{\cal C}[\mid a \rangle \,\mid f(a)\rangle]&=&0 \,\,otherwise.
\end{eqnarray}
Grover's loop of transformations is then defined by rotating a $\pi$ angle
the state $\mid a\rangle$ if ${\cal C}[\mid a \rangle \,\mid y^a\, mod\,
N\rangle]\,=\,1$ and doing nothing otherwise. Once we do that we do the
inversion about the average as in Grover's algorithm. At the end of ${\cal
O}(N)$ steps we will get:
\begin{equation}
\mid \eta
\rangle\,=\,\frac{1}{\sqrt{\tau}}\,\sum_{j\,=\,0}^{\tau\,-\,1}\,\mid\,1\,+\,jr\rangle
\end{equation}
where $\tau$ is the greatest integer less than $\frac{q\,-\,1}{r}$. So,
just doing three measurement operations over the state $\mid\eta\rangle$
one finds \cite{REVIEW}, with high probability, the period {\it r}. 
As discussed in the first part of this note the whole Grover's process is
unitary preserving constant the Fisher information function. In terms of
time it takes an exponential time of ${\cal O}(2^{\log N})$. The
difference between the {\bf fast} projection from (\ref{shori}) to
(\ref{shorf}) performed in Shor's algorithm and the {\bf adiabatic} slow
one using Grover's loop defined above is that in the adiabatic one the
complexity class is not changed and Fisher's function remains constant,
playing the classical role of entropy.
The quantum adiabatic algorithm using Grover's loop is certainly more
efficient than the classical one and very likely more robust with respect
to quantum decoherence problems than the faster Shor's algorithm.
Technologically is more feasible using for instance the recent
implementation of Grover's algorithm \cite {EXP1}, \cite {EXP2}.\\[.2cm]
{\bf \large 6.- }To finish we would like to suggest the following general
conjecture: \\[.2 cm]
{\it  Changes in complexity class should involve no conservation of
Fisher's information function and reciprocally constant Fisher's
information will not change the complexity class.}\\[.2 cm]
Our exercise also shows that the typical non unitary quantum projection
from (\ref{shori}) to (\ref{shorf}) used by Shor's algorithm can be, for
the practical purpposses of quantum computation, done using only unitary
transformations. The bill you have to paid for adiabaticity is longer time
and not change of complexity class.

\end{multicols}
\end{document}